\documentclass[prl,twocolumn,showpacs,superscriptaddress,fleqn,floatfix,footinbib,preprintnumbers]{revtex4-1}
\usepackage{graphicx}
\usepackage{amsmath}
\usepackage{amssymb}
\usepackage{stix}
\usepackage{braket}

\begin{document}

\title{Non-Adiabatic Vibrational Damping of Molecular Adsorbates:\\ 
Insights into Electronic Friction and the Role of Electronic Coherence}

\author{Simon P. Rittmeyer}
\email[Corresponding author: ]{simon.rittmeyer@tum.de}
\affiliation{Chair for Theoretical Chemistry and Catalysis Research Center, 
	Technische Universit\"at M\"unchen, Lichtenbergstr. 4, 85747 Garching, Germany}
\author{J\"org Meyer}
\affiliation{Leiden Institute of Chemistry, Gorlaeus Laboratories, Leiden 
	University, P.O. Box 9502, 2300 RA Leiden, The Netherlands}
\author{Karsten Reuter}
\affiliation{Chair for Theoretical Chemistry and Catalysis Research Center, 
	Technische Universit\"at M\"unchen, Lichtenbergstr. 4, 85747 Garching, Germany}

\begin{abstract}
We present a perturbation approach rooted in time-dependent density-functional theory to calculate electron hole (\emph{eh})-pair excitation spectra during the non-adiabatic vibrational damping of adsorbates on metal surfaces. Our analysis for the benchmark systems CO on Cu(100) and Pt(111) elucidates the surprisingly strong influence of rather short electronic coherence times. We demonstrate how in the limit of short electronic coherence times, as implicitly assumed in prevalent quantum nuclear theories for the vibrational lifetimes as well as electronic friction, band structure effects are washed out. Our results suggest that more accurate lifetime or chemicurrent-like experimental measurements could characterize the electronic coherence.
\end{abstract}

\pacs{82.65.+r, 
	34.50.Bw, 
	82.20.Gk  
	68.35.Ja  
}

\preprint{(Accepted for publication in Phys. Rev. Lett.)}

\maketitle

The tortuous ways in which kinetic and chemical energy is transferred between adsorbates and substrate atoms fundamentally govern the dynamics of surface chemical reactions, for instance in the context of heterogeneous catalysis or advanced deposition techniques. For metal substrates, the two main energy dissipation mechanisms in this regard are the adsorbate interaction with lattice vibrations, i.e., substrate phonons, and the excitation of electron hole (\emph{eh})-pairs. The latter are attributable to the non-adiabatic coupling of nuclear motion to the substrate electronic degrees of freedom and seem to be substantial in order to rationalize an increasing number of experimental findings \cite{Tully_JChemPhys_2012,Wodtke_ChemSocRev_2016}. Important steps towards an accurate, yet efficient first principles-based modeling of the energy uptake into phononic degrees of freedom have recently been taken \cite{Gross_PhysRevLett_2009, Meyer_Diss_2012, Meyer_AngewChemIntEd_2014, Nattino_JPhysChemLett_2016, Bukas_PhysRevLett_2016, Bukas_JChemPhys_2017, Kolb_JPhysChemLett_2017, Shakouri_JPhysChemLett_2017}. In contrast, the explicit description of \emph{eh}-pair excitations and corresponding non-adiabatic couplings directly from first principles still poses a formidable challenge.
	
In this regard, electronic friction theory (EFT) \cite{DAgliano_PhysRevB_1975, HeadGordon_JChemPhys_1995} has become a popular work horse to effectively capture the effects of such non-adiabatic energy loss on the adsorbate dynamics in a computationally convenient way \cite{Luntz_JChemPhys_2006, Juaristi_PhysRevLett_2008, BlancoRey_PhysRevLett_2014, Buenermann_Science_2015, Askerka_PhysRevLett_2016, Jiang_JPhysChemLett_2016, Rittmeyer_PhysRevLett_2016}. Inspired by vibrational lifetimes obtained via response theory \cite{Hellsing_PhysScripta_1984} or Fermi's Golden Rule in the nuclear system \cite{HeadGordon_JChemPhys_1992}, a Langevin equation for the nuclei emerges from a semi-classical picture implying complete electronic decoherence in terms of the Markov approximation \cite{HeadGordon_JChemPhys_1995}. This approach thus avoids an explicit propagation of the electron dynamics and concomitant ultrafast time scales by coarse-graining the effects into electronic friction forces linear in nuclear velocities. This enables an efficient combination even with density-functional theory (DFT) based {\em ab initio} molecular dynamics (AIMD) simulations on high-dimensional potential energy surfaces as required for surface dynamical studies \cite{BlancoRey_PhysRevLett_2014, Saalfrank_JChemPhys_2014, Novko_PhysRevB_2016}. 

Independent of the particular recipe employed to obtain the electronic friction coefficients  \cite{Hellsing_PhysScripta_1984, HeadGordon_JChemPhys_1995, Trail_PhysRevLett_2002, Juaristi_PhysRevLett_2008, Rittmeyer_PhysRevLett_2015, Askerka_PhysRevLett_2016}, however, the downside of the coarse-graining of the electron dynamics is that it precludes a more fundamental understanding of the underlying \emph{eh}-pair excitations. For instance, recent such calculations for non-adiabatic vibrational lifetimes of several small molecules \cite{Rittmeyer_PhysRevLett_2015, Askerka_PhysRevLett_2016, Maurer_PhysRevB_2016}, still do not elucidate the seemingly non-systematic trends for different adsorbate-substrate combinations \cite{Krishna_JChemPhys_2006, Forsblom_JChemPhys_2007, Maurer_PhysRevB_2016}.

Going beyond EFT approach is conceptually challenging, in particular without sacrificing a predictive-quality description of the metallic band structure at least on the DFT level. In principle, mixed quantum-classical dynamics in terms of Ehrenfest dynamics can provide access to \emph{eh}-pair excitation spectra \cite{Lindenblatt_PhysRevLett_2006,Grotemeyer_PhysRevLett_2014}. Notwithstanding, in the context of surface dynamical studies at extended metal surfaces this approach struggles with exceeding computational costs. In this letter we therefore pursue a perturbation approach rooted in time-dependent DFT (TD-DFT) that provides a computationally more appealing access to explicit \emph{eh}-pair excitation spectra. Aiming to scrutinize the confusing and inconclusive picture obtained from EFT as well as from experimental measurements, we revisit the vibrational damping of the CO stretch mode on Cu(100) and Pt(111). These represent two established benchmark systems for which accurate experimental lifetimes are available \cite{Morin_JChemPhys_1992,Beckerle_JChemPhys_1991} and energy loss into phononic degrees of freedom is commonly considered to be negligible \cite{Tully_JVacSciTech_1993,Saalfrank_ChemRev_2006,Arnolds_ProgSurfSci_2011}. In contrast to previous EFT-based work \cite{Rittmeyer_PhysRevLett_2015}, the non-adiabatic energy loss derived from our calculated \emph{eh}-pair excitation spectra differs significantly for both systems when electronic coherence is longer than five vibrational periods ($\approx80\,\mathrm{fs}$). In this case, as intuitively expected from the much higher density of states (DOS) in the vicinity of the Fermi level, we find a notably higher non-adiabatic energy dissipation rate for CO on Pt(111). In the limit of short electronic coherence times, i.e., approaching the Markov limit for the electronic degrees of freedom as relied upon in EFT, however, the detailed dependence on the metallic band structure is washed out.

Since the CO stretch mode is far above the substrate phonon continuum, the vibrational lifetime on a metal surface ($\hbar\omega_\mathrm{vib}\approx260\,\mathrm{meV}$) is commonly assumed to be one of few experimentally accessible observables that is governed by non-adiabatic energy dissipation \cite{Saalfrank_ChemRev_2006, Arnolds_ProgSurfSci_2011}. 
For both benchmark systems CO on Cu(100) and Pt(111) experimental reference lifetimes are around $2\,\mathrm{ps}$ \cite{Morin_JChemPhys_1992, Beckerle_JChemPhys_1991}. Even though these lifetimes could be nicely reproduced by EFT using different models for the friction coefficient \cite{Krishna_JChemPhys_2006, Forsblom_JChemPhys_2007, Rittmeyer_PhysRevLett_2015, Askerka_PhysRevLett_2016, Maurer_PhysRevB_2016}, the missing correlation of the vibrational lifetime with the underlying metal band structure remains an unresolved mystery \cite{Krishna_JChemPhys_2006, Forsblom_JChemPhys_2007, Maurer_PhysRevB_2016}. As a transition metal with an only partially filled $d$-band, platinum shows a significantly higher metallic density of states (DOS) in the energetically relevant region close to the Fermi level as compared to a coinage metal such as copper \cite{Hammer_SurfSci_1995}. Intuitively, it should thus allow for more \emph{eh}-pair excitations \cite{Note_LDFA_density}.
However, neither this nor the different adsorbate-induced DOS on both substrates seems to result in notable variations of the non-adiabatic vibrational lifetimes \cite{Forsblom_JChemPhys_2007}. 

There have been several attempts to deduce a more detailed understanding of the \emph{eh}-pair excitations behind frictional energy losses \cite{Trail_PhysRevLett_2002, Trail_JChemPhys_2003, Luntz_JChemPhys_124_2006} by connecting the latter to a forced oscillator model (FOM) \cite {Schoenhammer_1984}. In essence, the FOM describes electronic excitations in the substrate through a collection of independent harmonic oscillators driven by an external force of identical functional form but different strength \cite{Luntz_JChemPhys_124_2006}. It may thus be seen as a simple illustration of the ideas also underlying EFT \cite{Luntz_JChemPhys_124_2006}. Excitation spectra predicted by the FOM for the vibrational damping dynamics of CO on Cu(100) and Pt(111) are shown in Fig.~\ref{fig:spectra} below. Specifically, the hole excitation spectrum $P_{\mathrm{ex},h}(\epsilon_i)$ denotes the probability that created \emph{eh}-pair excitations involve the formation of a hole in the occupied energy level $\epsilon_i$, whereas the electron excitation spectrum $P_{\mathrm{ex},e}(\epsilon_j)$ shows this probability as a function of the unoccupied level $\epsilon_j$ that is filled with an electron. The symmetric sigmoid-like spectra stepped at the stretch vibrational frequency look exactly the same for both systems, as one would expect from their similar vibrational frequencies and friction coefficients. This functional form of the spectra is in fact a direct consequence of the motion pattern underlying the nuclear dynamics. It is essentially independent of the precise recipe used to obtain the electronic friction coefficients (cf.~supplemental material (SM) \cite{SI}), and it does not exhibit any correlation with the underlying metallic band structure \cite{Trail_PhysRevLett_2002}.

To scrutinize this picture we follow a perturbation approach to explicitly describe adsorbate-induced \emph{eh}-pair excitations in the metallic substrate. Originally developed for molecular scattering \cite{Timmer_PhysRevB_2009,Meyer_NewJPhys_2011}, we here present its straightforward extension to periodic motion. According to the Runge-Gross theorem \cite{Runge_PhysRevLett_1984} mapping to an effective single-particle picture described by $\hat{h}(t) = \hat{h}_0 + \hat{v}_\mathrm{pert}(t)$ is possible. Here, $\hat{h}_0$ is the unperturbed static Hamiltonian corresponding to the adsorbed molecule in its equilibrium geometry and $\hat{v}_\mathrm{pert}(t)$ describes the periodic time-dependent electronic perturbation potential exerted by the nuclear vibrational motion. The key-idea to arrive at a computationally tractable scheme that avoids an explicit time-dependent evaluation of the perturbation potential is to approximate the latter through a series of snapshots along the periodic molecular trajectory $\boldsymbol{Q}(t)$ around the equilibrium geometry $\boldsymbol{Q}_{0}$ such that
\begin{align}
	\hat{h}(t)&\approx \hat{h}_0 + \underbrace{\hat{v}_\mathrm{eff}(\boldsymbol{Q}(t)) - \hat{v}_\mathrm{eff}(\boldsymbol{Q}_0)}_{=\hat{v}_\mathrm{pert}(\boldsymbol{Q}(t))}\;, \label{eq:hamiltonian}
\end{align}
where $\hat{v}_\mathrm{eff}(\boldsymbol{Q})$ is the effective Kohn-Sham (KS) single particle potential at the respective snapshots. Each of these snapshots can in turn be treated within the framework of time-independent DFT, as the dynamic information has entirely been shifted into the time-dependence of $\boldsymbol{Q}(t)$. For the Hamiltonian defined in Eq.~(\ref{eq:hamiltonian}), first-order time-dependent perturbation theory then leads to the \emph{eh}-pair excitation spectrum at time $t_n=nT$, i.e., the probability to generate {\em eh}-pair excitations of energy $\epsilon$ after $n$ molecular oscillations with period $T$, as 
\begin{align}
	P_\mathrm{ex}(\epsilon;t_n) = \sum_{i,j}\left|\frac{\lambda_{ij}(t_n)}{\epsilon_j-\epsilon_i}\right|^2\delta\left(\epsilon - \epsilon_{ji} \right)\;. \label{eq:Pex}
\end{align}
Here, $\epsilon_{ji} = \epsilon_j-\epsilon_i$ is the energy difference between KS states $i,j$ of the unperturbed system and with eigenenergies $\epsilon_i, \epsilon_j$, while $\lambda_{ij}(t_n)$ is the corresponding transition matrix element. Separate hole $P_{\mathrm{ex},h}(\epsilon_i;t_n)$ and electron $P_{\mathrm{ex},e}(\epsilon_j;t_n)$ spectra are obtained by equations analogous to Eq. (\ref{eq:Pex}), in which the energy of initial and final state relative to the Fermi level rather than the energy difference $\epsilon_{ji}$ is considered \cite{Timmer_PhysRevB_2009, Meyer_NewJPhys_2011}. Integration by parts allows to calculate the matrix elements in Eq.~(\ref{eq:Pex}) for an $i\to j$ transition according to
\begin{align}
\lambda_{ij}(t_n) = 
\int_{0}^{t_n}\!\Braket{j|\nabla_{\boldsymbol{Q}}\hat{v}_\mathrm{pert}\left(\boldsymbol{Q}(t)\right)|i}\cdot\dot{\boldsymbol{Q}}(t)\,\operatorname{e}^{i\frac{\epsilon_{ji}}{\hbar}t}\,\mathrm{d}t\;.
\label{eq:lambda}
\end{align}
The integral limits are chosen such that the integration spans integer multiples of a vibrational period and the boundary terms conveniently vanish \cite{Meyer_NewJPhys_2011}. Further details on our evaluation of Eq.~(\ref{eq:lambda}) within the framework of 
semi-local DFT at the generalized-gradient level \cite{Clark_ZKristallogr_2005,Perdew_PhysRevLett_1996,*Perdew_PhysRevLett_1997} 
are presented in the SM \cite{SI}. Finally, the actual non-adiabatic energy loss is consistently evaluated as energy-weighted integral over the \emph{eh}-pair spectrum \cite{Timmer_PhysRevB_2009,Meyer_NewJPhys_2011} such that we can approximate the energy dissipation rate as
\begin{align}
	\gamma_\mathrm{\emph{eh}-pairs} =
	\frac{1}{t_n}
	\int_0^\infty \!\epsilon P_\mathrm{ex}(\epsilon;t_n)\,\mathrm{d}\epsilon\;. 
	\label{eq:gamma_eh}
\end{align}	

\begin{figure}[t]
	\includegraphics[]{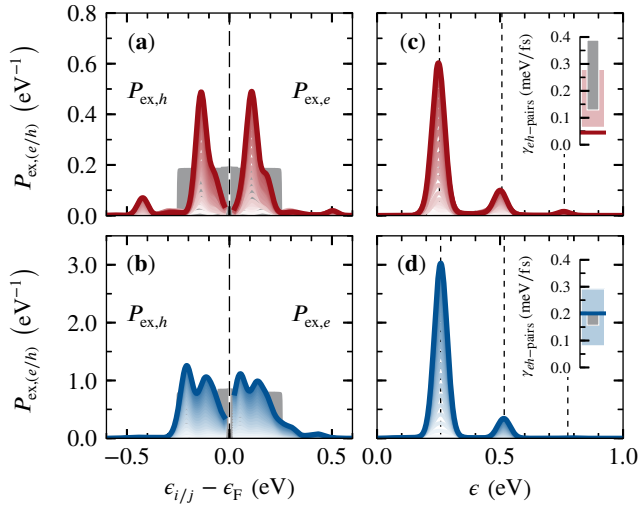}
	\caption{Excitation spectra for the non-adiabatic coupling of the stretch mode of CO on Cu(100) (upper panels, red) and Pt(111) (lower panels, blue) to the metallic continuum assuming infinite electronic coherence. The corresponding long-time limit is achieved by integrating Eq.~(\ref{eq:lambda}) over increasing times. The resulting spectral growth is indicated by lines of decreasing brightness, with the final bold line corresponding to an integration over $t_n =20T\approx 320\,\mathrm{fs}$. 
	Left panels (a) and (b) show electron and hole excitation spectra ($P_{\mathrm{ex},e}$ and $P_{\mathrm{ex},h}$) and additionally contain spectra calculated from the forced oscillator model (FOM, gray-shaded areas). Note that for direct comparability the FOM spectra are scaled to equal area with the respective explicitly evaluated spectra from perturbation theory. Right panels (c) and (d) show \emph{eh}-pair excitation spectra with peaks at multiples of the CO vibrational stretch frequency (dotted vertical lines). The insets depict the corresponding non-adiabatic energy loss rates as compared to experiment (gray-shaded area indicating the reported error bars) \cite{Morin_JChemPhys_1992, Beckerle_JChemPhys_1991} and to the range of values obtained from EFT (red/blue-shaded range) \cite{Rittmeyer_PhysRevLett_2015}.}
	\label{fig:spectra}
\end{figure}

We first note the striking similarity between Eq.~(\ref{eq:lambda}) and the matrix elements required in the context of an orbital-dependent formulation of the electronic friction tensor (orbital-dependent friction, ODF) \cite{Hellsing_PhysScripta_1984, HeadGordon_JChemPhys_1995, Trail_JChemPhys_2003, Askerka_PhysRevLett_2016, Maurer_PhysRevB_2016}. This is not surprising as---in contrast to the popular purely embedding density-based local density friction approximation (LDFA) \cite{Echenique_SolidStateCommun_1981, Echenique_PhysRevA_1986, Li_PhysRevLett_1992, Juaristi_PhysRevLett_2008}---ODF is also based on a time-dependent perturbation treatment of the \emph{eh}-pair excitations. Alike \cite{Lorente_FaradayDiscuss_2000,Maurer_PhysRevB_2016}, we consider only inter-$\mathbf{k}$ and thus intraband transitions here in order to mimic the coherent vibrational excitation in experiments. However, the subtle difference is that our approach targets effects of the non-adiabatic coupling on the electronic rather than the nuclear degrees of freedom, allowing to explicitly evaluate excitation spectra and to follow the coherent time evolution of the electronic excitations for varying propagation times (cf.~SM \cite{SI}). In contrast, ODF does not explicitly describe electronic excitations as a function neither of energy nor of time, but condenses the matrix elements at the Fermi level into a friction coefficient which then acts as effective energy sink for the nuclear motion. This implies a constant coupling in the region energetically relevant to the \emph{eh}-pair excitations \cite{HeadGordon_JChemPhys_1995}. Unlike what is assumed in EFT-equivalent lifetime theory \cite{HeadGordon_JChemPhys_1992, Maurer_PhysRevB_2016} our perturbation operator further deviates from a purely harmonic perturbation due to the consideration of spatially varying transition matrix elements $\lambda_{ij}(t)$. Correspondingly, the \emph{eh}-pair excitation spectra calculated using an unperturbed trajectory for the long time limit, which is in practice already reached after $t_n=5T = 80\,\mathrm{fs}$ for both systems, also exhibit higher-frequency components, cf. Fig.~\ref{fig:spectra}. Such overtones that correspond to excitations of multiple quanta of the molecular stretch frequency arise in the present formalism as a consequence of the simultaneous consideration of infinite electronic coherence and a perturbation that constantly drives \emph{eh}-pair excitations without being affected by the latter.

Integrating the \emph{eh}-pair excitation spectra via Eq.~(\ref{eq:gamma_eh}) allows to compare the resulting non-adiabatic energy dissipation rate to the one predicted by EFT, both within the LDFA and ODF (cf.~SM for further details \cite{SI}). First of all, we note that in both theories these rates are very small---less than $1.5\%$ of the vibrational quantum is dissipated per vibrational period---suggesting that first-order perturbation theory should be valid. As further apparent from Fig.~\ref{fig:spectra} our results for CO on Pt(111) fall within the corresponding range spanned by EFT. In contrast, a lower rate is obtained for CO on Cu(100). Consequently, a significantly different non-adiabatic energy dissipation arises for the two systems, exactly as one would have expected on the basis of their metallic band structure. This difference results irrespective of the higher-frequency overtones and is robust, even if the latter are excluded from the integration. The underlying electron and hole excitation spectra shown in Fig.~\ref{fig:spectra} rationalizes this varying agreement between the two theories for both metals. On Pt(111) the most dominant spectral contributions evenly originate from excitations within $250\,\mathrm{meV}$ around the Fermi level. Overall, this yields a spectral shape that largely resembles the sigmoid shape of the friction-inspired FOM, and---without being imposed---justifies the constant-coupling approximation underlying EFT, which assumes an equal excitation probability of states in the vicinity of the Fermi level and yields the decoherent (Markov) limit \cite{HeadGordon_JChemPhys_1995}. In contrast, on Cu(100) the vibrational motion of the CO molecule specifically triggers excitations from and to a very narrow region of initial and final states. The corresponding spectra thus exhibit prominent peaks at about $\pm130\,\mathrm{meV}$, but only comparably little contributions directly around the Fermi level as expected within the FOM and EFT. As such, the electron and hole excitation spectra for the two substrates as obtained in the perturbation approach directly reflect their very different band structure: An abundance of available \emph{eh}-pairs resulting in a broad range of excitations on the transition metal as opposed to a distinct coupling between a limited number of states on the coinage metal. Simultaneously, the occurrence and location of the pronounced excitation peaks for CO on Cu(100) rationalize for instance the very large broadening used to obtain ODF tensor elements for this system \cite{Askerka_PhysRevLett_2016,Maurer_PhysRevB_2016,Note_broadening}

Within the long-time limit the present theory thus sketches an intuitive picture of the {\em eh}-pair excitations governing the vibrational damping. On the other hand, the predicted difference in non-adiabatic energy dissipation rates for the two systems is not reflected in the measured lifetimes, cf. Fig.~\ref{fig:spectra}. This calls to scrutinize key assumptions entering the perturbation approach in this long-time limit. Notably, these first concern a perturbation operator that is unaffected by the excitations it triggers which allows to cast the respective time evolution into an unperturbed trajectory in the nuclear subspace. Second, addressing the electronic subspace, we have so far assumed the persistence of electronic phase coherences as induced through the perturbation for infinite times. We can estimate the consequences of the prior approximation by considering an exponential damping of the (nuclear) vibrational amplitude due to the ongoing energy dissipation. Adding a corresponding exponential damping factor in Eq.~(\ref{eq:lambda}) simply yields long-time spectra that are convoluted with a Lorentzian of width proportional to the damping, but at otherwise unchanged spectral weight. As such, this approximation unlikely affects the relative dissipation rate of the two systems. 

\begin{figure}[t]
	\includegraphics[]{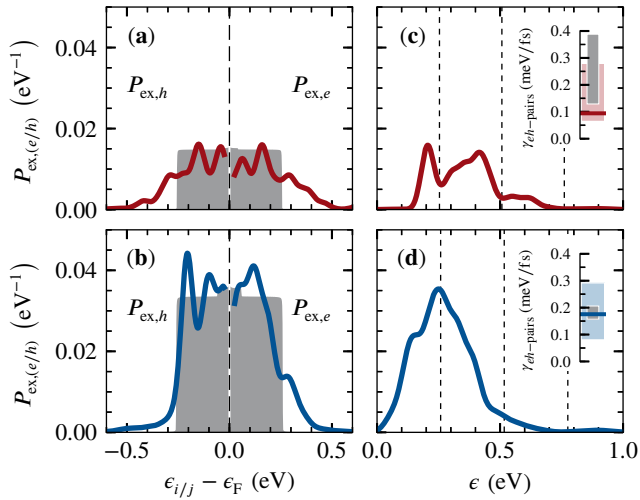}
	\caption{Same as Fig.~\ref{fig:spectra} but for integration over one vibrational period only to mimic the effect of a finite electronic phase coherence. Please note that opposed to the energy dissipation rates in the insets, the actual spectra are not normalized with respect to the integration time and thus show significantly lower excitation probabilities as compared to Fig.~\ref{fig:spectra}.}
	\label{fig:spectra_short}
\end{figure}

In contrast, the assumption of infinite electronic coherence allows only {\em eh}-pair excitations that are resonant with the vibrational perturbation to accumulate appreciable transition probabilities. Shorter coherence times will consequently have immediate impact on the spectral shape. In the present perturbation approach we can easily assess this by integrating Eq.~(\ref{eq:lambda}) over one vibrational period ($T\approx 16\,\mathrm{fs}$) only, thus assuming that all phase information is reset after this characteristic period of time.
The spectra calculated for such pseudo-Markovian systems are shown in Fig.~\ref{fig:spectra_short}. Interestingly, even though a continuum of off-resonant \emph{eh}-pair excitations is possible now for both systems, this has little effect on the electron and hole excitation spectra, as well as on the energy dissipation rate in the case of Pt(111). This is consistent with the constant coupling approximation being much better fulfilled for this system even in the long time limit. In contrast, both $P_{\mathrm{ex},h}(\epsilon_i)$ and $P_{\mathrm{ex},e}(\epsilon_j)$ are strongly modulated in the case of Cu(100) and now resemble much more the sigmoid-like shape of the FOM. Consistently, the energy dissipation rate rises and becomes more compatible with EFT-based results. In consequence, the relative difference in the energy dissipation rate of the two systems decreases and gets closer to the experimental findings.

As such, shorter electronic coherence times could indeed be one reason why the largely different metallic band structure of the two systems does not markedly show up in the vibrational lifetimes. The concomitant possibility to generate off-resonant excitations provides so many individual excitation channels that the detailed density of initial and final states averages out. Incidentally, this then justifies the constant-coupling approximation fundamentally relied on in EFT \cite{HeadGordon_JChemPhys_1995} and could rationalize the good agreement of the vibrational lifetimes obtained with this approach. Notwithstanding, even in the present limit of a finite electronic coherence time over one vibrational period only, the energy dissipation rate on Pt(111) remains about double as large as the one on Cu(100). Especially due to the large error bars for the measured vibrational lifetime of CO on Cu(100) \cite{Morin_JChemPhys_1992}, it is unclear if this small difference is compatible with the experimental data. New high-precision measurements would thus be most valuable in this respect. Likewise, our calculated spectra suggest that chemicurrent-like measurements as pioneered by Nienhaus \emph{et al.} \cite{Nienhaus_PhysRevLett_1999} could allow to distinguish the two limits for the electronic coherence times. Should such measurements disconfirm the present theoretical prediction, we speculate that other hitherto disregarded dissipation channels have a notable contribution to the measured lifetimes, thus questioning the notion of purely non-adiabatic vibrational energy losses. Likely candidates to this end are small, but non-negligible couplings to other adsorbate vibrational modes or the substrate phononic system.

In conclusion, we have extended a numerically efficient TD-DFT based perturbation approach to evaluate explicit \emph{eh}-pair excitation spectra for the vibrational damping of high-frequency adsorbate modes---a phenomenon that stimulated the popular EFT approach and is still of crucial importance for its justification. This framework allows us to study the detailed effect of different models for the perturbation operator on the electronic subsystem and thus to explicitly test assumptions about the electronic dynamics entering the EFT formalism directly at its roots. Our analysis shows an unexpectedly large influence of electronic coherence which allows to rationalize the hitherto enigmatic similarity of measured vibrational lifetimes of CO on Pt(111) and Cu(100). In the limit of infinite electronic coherence, specific state-to-state coupling dominates a smaller non-adiabatic energy dissipation for CO on Cu(100), as one would intuitively expect from the coinage metal band structure. It requires a finite electronic coherence time to open up off-resonant {\em eh}-pair excitation channels that raise the dissipation rate to similar levels as found for CO on Pt(111). The resulting multitude of excitation channels washes out the differences in the underlying transition and coinage metal band structure and leads to unstructured electron and hole excitation spectra. Such a spectral shape can in principle be measured experimentally, and thus constitutes a direct access to most fundamental electronic coherence times. Moreover, the resulting constant coupling is implicitly assumed in EFT, which might explain the good performance of this effective theory in reproducing the experimental lifetimes. Intriguingly, a small relative difference in dissipation rates between the two benchmark systems remains even in the limit of short electronic coherence times. Most accurately determined by new high-precision experiments this difference is thus a sensitive measure of contributions of hitherto disregarded dissipation mechanisms like vibrational mode coupling or residual coupling to phononic degrees of freedom, and thus allows to ultimately shed light into fundamental questions such as the actual degree of non-adiabaticity of the vibrational energy decay.

\begin{acknowledgments}
S.P.R. acknowledges support of the Technische Universit\"at M\"unchen - Institute for Advanced Study, funded by the German Excellence Initiative and the European Union Seventh Framework Programme under grant agreement No. 291763, and financial support in terms of an STSM within COST Action MP1306 (EUSpec). J.M. is grateful for financial support from the Netherlands Organisation for Scientific Research (NWO) under Vidi Grant No. 723.014.009.
\end{acknowledgments}

\bibliography{literature,footnotes}
\end{document}